\documentclass{article}
\usepackage{amssymb}
\usepackage{amsmath}
\usepackage{graphicx}

\begin{document}

\begin{center}
\bigskip \textbf{ON\ GENERALIZED WORMHOLE IN THE EDDINGTON INSPIRED BORN
INFELD (EiBI) GRAVITY}

Amarjit Tamang$^{1,a}$, Alexander A. Potapov$^{2,b}$, Regina Lukmanova$%
^{3,c} $,

Ramil Izmailov$^{3,d}$ and Kamal K. Nandi$^{2,3,4,e}$

$\bigskip $

$^{1}$Department of Mathematics, Darjeeling Government College, Richmond
Hill, Darjeeling 734104, WB, India

$^{2}$Department of Physics \& Astronomy, Bashkir State University,
Sterlitamak Campus, Sterlitamak 453103, RB, Russia

$^{3}$Zel'dovich International Center for Astrophysics, M. Akmullah Bashkir
State Pedagogical University, Ufa 450000, RB, Russia \\[0pt]
$^{4}$Department of Mathematics, University of North Bengal, Siliguri
734013, WB, India \\[0pt]

\bigskip

$^{a}$Email: amarjit.tamang1986@gmail.com

$^{b}$Email: potapovaa@mail.ru

$^{c}$Email: mira789@mail.ru

$^{d}$Email: izmailov.ramil@gmail.com

$^{e}$Email:kamalnandi1952@yahoo.co.in

-----------------------------------------------------

\textbf{Abstract}
\end{center}

In this paper, we wish to investigate certain observable effects in the
recently obtained wormhole solution of the EiBI theory, which generalizes
the zero mass Ellis-Bronnikov wormhole of general relativity. The solutions
of EiBI theory contain an extra parameter $\kappa $ having the inverse
dimension of the cosmological constant $\Lambda $, and is expected to modify
various general relativistic observables such as the masses of wormhole
mouths, tidal forces and light deflection. A remarkable result is that a
non-zero $\kappa $ could prevent the tidal forces in the geodesic
orthonormal frame from becoming arbitrarily large near a small throat radius
($r_{0}\sim 0$) \textit{contrary} to what happens near a small Schwarzschild
horizon radius ($M\sim 0$). The role of $\kappa $ in the flare-out and
energy conditions is also analysed, which reveals that the energy conditions
are violated. We show that the exotic matter in the EiBI wormhole cannot be
interpreted as phantom ($\omega =\frac{p_{r}}{\rho }<-1$) or ghost field $%
\phi $ of general relativity due to the fact that both $\rho $ and $p_{r}$
are negative for all $\kappa $.


\section{%
\baselineskip=4ex%
\textbf{Introduction}}

One of the fundamental discoveries in astrophysics in recent times is that
the universe is currently accelerating [1,2]. A possible explanation for the
late-time cosmic acceleration could be due to the infra-red modifications
[3] of Einstein's General Relativity (GR). Such alternative theories of
gravity involve more general combinations of curvature invariants than the
pure Einstein-Hilbert term. One such modified theory is the
Eddington-inspired Born-Infeld (EiBI) gravity. What is this EiBI gravity? It
is a prototype of theories that could be termed as the "gravitational avatar
of non-linear electrodynamics" [4].

To be more specific, note that Eddington's original gravitational action is
incomplete in the sense that it does not contain matter. Ba\~{n}ados and
Ferreira [5] resurrected Eddington's proposal for the gravitational action
in the presence of cosmological constant extending it to include matter
fields in the form of a Born-Infeld like structure [6] of non-linear
electrodynamics. The outcome is the modern form of EiBI gravity, which
provides an alternative theory of the Big Bang with a novel, non-singular
description of the Universe. The EiBI\ model is currently extensively
applied in the literature to many other astrophysical scenarios such as the
solar system, structure of neutron stars or dark matter etc [7-14].
Astrophysical scenarios today also include wormholes as an integral part,
and we shall be dealing with one such solution here.

The solutions of the EiBI theory contain an extra parameter $\kappa $ having
the inverse dimension of the cosmological constant $\Lambda $, that is,
(length)$^{2}$. The theory is ideologically relatively new and very
different from GR, except in the limit $\kappa \rightarrow 0$. Thus, the
true EiBI theory must always have $\kappa \neq 0$, and this parameter is
expected to modify different GR physical observables. In the same spirit, we
wish to investigate the effect of $\kappa $ on the observable quantities
associated with a wormhole in EiBI theory. Such a wormhole has in fact been
recently derived by Harko \textit{et al.} [15], which could be regarded as a 
$\kappa \neq 0$ generalization of the original "zero total mass"
Ellis-Bronnikov (EB) wormhole of the Einstein minimally coupled scalar field
theory with a negative kinetic term\footnote{%
Recall that the 1973 Ellis "drainhole" solution [16] was independently
discovered also by Bronnikov [17]. The term "wormhole" was seemingly not in
vogue in 1973. Hence our current nomenclature EB wormhole, which belongs to
general relativity, hence corresponds to $\kappa =0$. The EiBI wormhole
derived in [15] can be called its EiBI generalization due to the presence of
the parameter $\kappa \neq 0$.}. Assuming that the EB wormhole has a
standard coordinate throat radius $r_{0}$, what we mean by zero total mass
here is that the individual masses in suitable units of the two mouths ($%
+r_{0}/2$ and $-r_{0}/2$) add exactly to zero, when $\kappa =0$. The new
generalized wormhole ($\kappa \neq 0$) derived by Harko \textit{et al.} [15]
is being extensively cited in the literature [18]. Thus, it is of interest
to find out what corrections $\kappa $ contribute to the observables of the
zero mass general relativistic EB wormhole.

The purpose of the present article is to derive several useful results that
can be stated as follows: (i) The zero total mass behavior is preserved even
when $\kappa \neq 0$. (ii) A non-zero $\kappa $ prevents the tidal forces in
the geodesic orthonormal frame from becoming arbitrarily large near $%
r_{0}\sim 0$, \textit{contrary} to what happens near a small Schwarzschild
horizon radius, $M\sim 0$. (iii) A non-zero $\kappa $ also influences light
bending, which provides a possibility to estimate $\kappa $ through
gravitational lensing observations. (iv) A non-zero $\kappa $ has a role in
the flare-out and energy conditions. (v) Finally, in the Appendix, we point
out the reasons why one cannot interpret the EiBI exotic matter either as
phantom or as ghost field considered in GR.

In Sec.2, we give a brief outline of the EiBI gravity to make the topic more
transparent. Then, in Sec.3, we briefly describe the wormhole under
investigation and calculate the masses of its two mouths. After a brief
review of tidal forces in a Lorentz boosted frame in Sec.4, we calculate in
Sec.5 the excess tidal forces experienced by a traveler in geodesic motion
near the throat of the wormhole. We devote Sec.6 to a discussion of the role
of $\kappa $ in the flare-out and energy conditions. In Sec.7, we calculate
the effect of $\kappa \ $on the bending of light passing by the positive
mass mouth. The final section (Sec.8) concludes the paper, followed by an
Appendix. We take units so that $G=1$, $c=1$.

\section{\textbf{\ EiBI field equations}}

In 1924, Eddington [19] suggested that at least in free, de-Sitter space,
the fundamental dynamical variable should be the affine connection $\Gamma $
and proposed a gravitational action $S_{\text{Edd}}=2\kappa \int d^{4}x\sqrt{%
\det \left\vert R_{\mu \nu }\left( \Gamma \right) \right\vert }$, where $%
\kappa $ is a constant with inverse dimension of $\Lambda $. Varying the
affine connection $\Gamma $, one obtains the field equations $\nabla
_{\alpha }\left( 2\kappa \sqrt{\left\vert R\right\vert }R^{\mu \nu }\right)
=0$, where $\nabla _{\alpha }$ is the covariant derivative defined by $%
\Gamma $ and $R^{\mu \nu }$ is the contravariant Ricci tensor. Eddington's
field equations can be solved if we define a new tensor $q_{\mu \nu }$ such
that $\nabla _{\alpha }\left( \sqrt{\left\vert q\right\vert }q^{\mu \nu
}\right) =0$. The field equations then become $2\kappa \sqrt{\left\vert
R\right\vert }R^{\mu \nu }=\sqrt{\left\vert q\right\vert }q^{\mu \nu }$,
which reduce to Einstein-de Sitter field equations if we identify $g_{\mu
\nu }=q_{\mu \nu }$ and $\kappa =\Lambda ^{-1}$. Thus Eddington's proposal
is a good motivation for building a more general action alternative to
Einstein's gravity. However, Eddington's theory does not include matter.
Therefore, Ba\~{n}ados and Ferreira [5] included matter, a metric $g_{\mu
\nu }$, a Born-Infeld [6] like structure replacing the pure affine Eddington
action by a new action that gave birth to what is now called EiBI theory in
the literature (for details, see [5]).

We shall focus on the EiBI theory embodied in the Ba\~{n}ados-Ferreira
action [5] given by\footnote{%
The action was first proposed by Vollick [20], but the matter fields were
introduced in a non-conventional way inside the square root, unlike in (1).} 
\begin{equation}
S_{\text{EiBI}}=\frac{1}{16\pi }\frac{2}{\kappa }\int d^{4}x\left[ \sqrt{%
\det \left\vert g_{\mu \nu }+\kappa R_{\mu \nu }(\Gamma )\right\vert }%
-\lambda \sqrt{\det \left\vert g_{\mu \nu }\right\vert }\right] +S_{\text{%
matter}}\left[ g,\Psi _{\text{matter}}\right]
\end{equation}%
where $\Psi _{\text{matter}}$ is a generic matter field, $\sqrt{\det
\left\vert g_{\mu \nu }+\kappa R_{\mu \nu }(\Gamma )\right\vert }$ is a
Born-Infeld like structure [6], $\lambda $ is a dimensionless parameter, $%
g_{\mu \nu }$ is the physical metric tensor, $R_{\mu \nu }(\Gamma )$ is the
symmetric part of the Ricci tensor built solely from the connection $\Gamma $%
, yet undefined. For small values of $\kappa R$, the action (1) reproduces
the Einstein-Hilbert action with a constant $\frac{\lambda -1}{\kappa }$,
identified as the cosmological constant (This will be more transparent from
the expansion of the field equations below): 
\begin{equation}
\Lambda =\frac{\lambda -1}{\kappa }.
\end{equation}%
For large values of $\kappa R$, the action approximates to matter-free
Eddington action $S_{\text{Edd}}$. To ensure asymptotic flatness of
solutions in the EiBI theory ($\kappa \neq 0$), one must have $\Lambda =0$,
which in turn would entail $\lambda =1$ from Eq.(2).

The field equations are based on a Palatini-type formulation, where the
metric tensor $g_{\mu \nu }$ and the connection $\Gamma $ are the two
independent dynamical variables that are varied in the action (1). Varying
with respect to $g_{\mu \nu }$, one obtains ($\left\vert X\right\vert $
denotes $\det \left\vert X_{\mu \nu }\right\vert $):%
\begin{equation}
\frac{\sqrt{\left\vert g+\kappa R\right\vert }}{\sqrt{\left\vert
g\right\vert }}\left[ \left( g+\kappa R\right) ^{-1}\right] ^{\mu \nu
}-\lambda g^{\mu \nu }=-8\pi \kappa T^{\mu \nu }\text{,}
\end{equation}%
where the usual stress tensor $T^{\mu \nu }$ is raised or lowered with $%
g_{\mu \nu }$. \ The field Eq.(3) expands as [5]: $R_{\mu \nu }\simeq \left( 
\frac{\lambda -1}{\kappa }\right) g_{\mu \nu }+T_{\mu \nu }-\frac{1}{2}%
g_{\mu \nu }T+\kappa \left[ S_{\mu \nu }-\frac{1}{4}g_{\mu \nu }S\right] ,$
where $S_{\mu \nu }=T_{\mu }^{\alpha }T_{\alpha \nu }-\frac{1}{2}TT_{\mu \nu
}$ and $\frac{\lambda -1}{\kappa }$ can be identified with $\Lambda $. Note
that Einstein's GR is recovered as $\kappa \rightarrow 0$.

The variation with respect to $\Gamma $ can be simplified by introducing an
auxiliary metric $q_{\mu \nu }$ compatible with $\Gamma $ defined by $\Gamma
_{\beta \gamma }^{\alpha }\equiv \frac{1}{2}q^{\alpha \sigma }\left[
\partial _{\gamma }q_{\sigma \beta }\text{ }+\partial _{\beta }q_{\sigma
\gamma }-\partial _{\sigma }q_{\beta \gamma }\right] $ so that the equation
of motion becomes\footnote{%
Alternatively, variation with respect to the connection $\Gamma $ leads to
the corresponding field equations. By defining $q_{\mu \nu }:=g_{\mu \nu
}+\kappa R_{\mu \nu }$ [Eq.(4)], after some manipulations, the field
equations take the form $\Gamma _{\beta \gamma }^{\alpha }=\frac{1}{2}%
q^{\alpha \sigma }\left[ \partial _{\gamma }q_{\sigma \beta }\text{ }%
+\partial _{\beta }q_{\sigma \gamma }-\partial _{\sigma }q_{\beta \gamma }%
\right] $ (see Ref.[12] for details). So, Eq.(4) and this form are exactly
equivalent field equations. This explains the genesis of the auxiliary
metric $q_{\mu \nu }$: It is coming from the variation of the dynamical
variable $\Gamma $. Of course, Eq.(4) is more illuminating, and we take it.
Only in vacuum $q_{\mu \nu }=g_{\mu \nu }$, but \textit{inside} matter they
are different. This is the essence of the EiBI theory.} 
\begin{equation}
q_{\mu \nu }=g_{\mu \nu }+\kappa R_{\mu \nu }.
\end{equation}%
Ba\~{n}ados and Ferreira [5] restricted their analysis to cases, where
matter couples \textit{only} to the metric $g_{\mu \nu }$ determining the
geodesic equation $\nabla _{\mu }T_{(g)}^{\mu \nu }=0$ but coupling to $%
\Gamma (q)$ may arise due to quantum gravitational corrections. Since the
auxiliary metric $q_{\mu \nu }$ is connected $g_{\mu \nu }$, EiBI does not
introduce any extra degree of freedom.

Thus, only the metric $g_{\mu \nu }$ is of physical interest as far as
gravitational observables are concerned. Combining (3) and (4), one finds%
\begin{equation}
\sqrt{\left\vert q\right\vert }q^{\mu \nu }=\lambda \sqrt{\left\vert
g\right\vert }g^{\mu \nu }-8\pi \kappa \sqrt{\left\vert g\right\vert }T^{\mu
\nu },
\end{equation}%
where $q^{\mu \nu }$and $g^{\mu \nu }$ are the matrix inverses of $q_{\mu
\nu }$ and $g_{\mu \nu }$ respectively. Eqs. (4) and (5) provide the
complete set of general EiBI field equations for arbitrary $\lambda $. In
vacuum ($T_{\mu \nu }=0$), $g_{\mu \nu }=q_{\mu \nu }$, $\Gamma =\Gamma (g)$
and hence EiBI and GR are completely equivalent.

For the specific case of asymptotically flat solutions, $\lambda =1$, and
hence Eq.(5) can be rewritten as%
\begin{equation}
q^{\mu \nu }=\tau \left( g^{\mu \nu }-8\pi \kappa T^{\mu \nu }\right) ,
\end{equation}%
where%
\begin{equation}
\tau =\sqrt{\frac{\left\vert g\right\vert }{\left\vert q\right\vert }}.
\end{equation}%
Harko \textit{et al.} [15] further simplified the Eqs.(4) and (6) combining
them into a form that looks much more familiar:%
\begin{eqnarray}
R_{\nu }^{\mu } &=&8\pi S_{\nu }^{\mu }, \\
S_{\nu }^{\mu } &=&\tau T_{\nu }^{\mu }-\left( \frac{1-\tau }{8\pi \kappa }+%
\frac{\tau }{2}T\right) \delta _{\nu }^{\mu },
\end{eqnarray}%
where $R_{\nu }^{\mu }=q^{\mu \sigma }R_{\sigma \nu }$, $R=R_{\mu }^{\mu }$,
and $T_{\nu }^{\mu }=T^{\mu \sigma }g_{\sigma \nu }$, $T=T_{\mu }^{\mu }$.
Note the roles of $q$ and $g$ metrics $-$ the Ricci tensor on the left hand
side of Eq.(8) is raised or lowered with $q$, while the right\ hand side is
done with the metric $g$.

\section{\textbf{Wormhole solution : Masses of the two mouths}}

The wormhole solution is derived in [15] by solving Eqs.(8) under certain
restrictive conditions such as spherical symmetry and asymptotic flatness,
the latter requiring $\lambda =1$. These assumptions of course limit the
applicability of EiBI theory but make the problem at hand much simpler to
handle. One spin-off is that the description of the physical behavior of the
wormhole is now controlled by the only remaining parameter $\kappa $. The
physical metric $g_{\mu \nu }$ and the auxiliary metric $q_{\mu \nu }$
respectively are taken as%
\begin{eqnarray}
g_{\mu \nu }dx^{\mu }dx^{\nu } &=&-e^{\nu (r)}dt^{2}+e^{\sigma
(r)}dr^{2}+f(r)d\Omega ^{2}, \\
q_{\mu \nu }dx^{\mu }dx^{\nu } &=&-e^{\beta (r)}dt^{2}+e^{\alpha
(r)}dr^{2}+r^{2}d\Omega ^{2}.
\end{eqnarray}%
The wormhole is assumed to be threaded by anisotropic matter described by
the stress tensor $T^{\mu \nu }=p_{t}g^{\mu \nu }+(p_{t}+\rho )U^{\mu
}U^{\nu }+(p_{r}-p_{t})\chi ^{\mu }\chi ^{\nu }$, where $\chi ^{\mu }$ is
the unique spacelike vector in the radial direction, $\chi ^{\mu
}=e^{-\sigma (r)/2}\delta _{r}^{\mu }$, $p_{r}$ is the radial pressure, $%
p_{t}$ is the tranverse pressure, $\rho $ is the energy density, $U^{\mu }$
is the four velocity such that $g_{\mu \nu }U^{\mu }U^{\nu }=-1$. Since
geodesics are determined by the metric $g_{\mu \nu }$, all observable
effects connected to geodesics such as light deflection or tidal forces
should be calculated only in the physical metric $g_{\mu \nu }$.

Note that $\tau $ of Eq.(7) can be obtained from $T_{\nu }^{\mu }$ through
the expression $\tau =\left\vert \delta _{\nu }^{\mu }-8\pi \kappa T_{\nu
}^{\mu }\right\vert ^{-1/2}$, which in turn can be expressed in terms of
stress quantities 
\begin{equation}
\tau =\left[ \left( 1+8\pi \kappa \rho \right) (1-8\pi \kappa p_{r})(1-8\pi
\kappa p_{t})^{2}\right] ^{-1/2}.
\end{equation}%
The above form suggests arbitrary functions $a$, $b$ and $c$ defined by 
\begin{eqnarray}
a(r) &=&\sqrt{1+8\pi \kappa \rho }, \\
b(r) &=&\sqrt{1-8\pi \kappa p_{r}}, \\
c(r) &=&\sqrt{1-8\pi \kappa p_{t}},
\end{eqnarray}%
that help one write the components of the field Eqs.(8) in manageable forms
that finally yield%
\begin{equation}
e^{\beta }=e^{\nu }\frac{c^{2}}{a^{2}}\text{, }e^{\alpha }=e^{\sigma
}a^{2}c^{2}\text{, }f=\frac{r^{2}}{ab}\text{.}
\end{equation}%
The specific wormhole solution obtained by Harko \textit{et al.} [15] is
based on simplifying assumptions that%
\begin{equation}
a(r)b(r)=1,\beta =0.
\end{equation}%
Then the reduced system of field Eqs.(8) yield%
\begin{equation}
e^{\alpha }=1-\frac{r_{0}^{2}}{r^{2}}\text{, }a^{4}=\frac{1}{1+2\kappa
r_{0}^{2}/r^{4}}\text{, }c^{2}=a^{2}.
\end{equation}%
These, together with Eqs.(16), lead to 
\begin{eqnarray}
q_{\mu \nu } &:&e^{\beta (r)}=1\text{, }e^{\alpha (r)}=\frac{1}{%
1-r_{0}^{2}/r^{2}},  \notag \\
g_{\mu \nu } &:&e^{\nu (r)}=1\text{, }e^{\sigma (r)}=\frac{1+2\kappa
r_{0}^{2}/r^{4}}{1-r_{0}^{2}/r^{2}},
\end{eqnarray}%
where $r_{0}$ is an arbitrary constant. Hence we have the metric $g_{\mu \nu
}$ given by (19), viz., 
\begin{equation}
g_{\mu \nu }dx^{\mu }dx^{\nu }=-dt^{2}+\left( \frac{1+2\kappa r_{0}^{2}/r^{4}%
}{1-r_{0}^{2}/r^{2}}\right) dr^{2}+r^{2}[d\theta ^{2}+\sin ^{2}\theta
d\varphi ^{2}],
\end{equation}%
The metric (20) is a symmetric, twice asymptotically flat regular wormhole
having asymptotic masses on either side of the throat, where $r_{0}$ has the
meaning that it is the standard coordinate throat radius $r_{\text{th}}=r_{0}
$, $r_{0}<r<+\infty $. In the limit $\kappa \rightarrow 0$, one recovers the
massless EB wormhole of GR [16,17]. 

To obtain the asymptotic masses, one needs to cover \textit{both} sides of
the wormhole by a single regular chart defined by%
\begin{equation}
r^{2}=\ell ^{2}+r_{0}^{2},
\end{equation}%
which is dictated by dimensional considerations, so the throat is now
appearing at $\ell _{\text{th}}=0$. Then the charts covering individual
sides respectively are $-\infty <\ell \leq 0$ and $0\leq \ell <+\infty $,
both meeting at the throat. Further, the structure of EiBI theory provides
an energy density that can be obtained from Eqs.(13) and (18) as 
\begin{equation}
\rho (r)=\frac{1}{8\pi \kappa }\left[ \frac{1}{\sqrt{1+2\kappa
r_{0}^{2}/r^{4}}}-1\right] ,
\end{equation}%
and the pressures from Eqs.(14), (15) and (18)%
\begin{equation}
p_{r}(r)=\frac{\rho (r)}{1+8\pi \kappa \rho (r)},\text{ }p_{t}=-\rho .
\end{equation}%
Fig.1 shows that $\rho (r)<0$, $p_{r}(r)<0$ for all values of $r$ and for
all values of $\kappa $ positive or negative. From Eq.(22), we can obtain
masses on individual sides using the prescription\footnote{%
In curved space with the metric (10), the volume measure contains $e^{\sigma
/2}$, while the measure "$4\pi r^{2}dr$" below follows the one in Ref.[21a]
already used for wormholes. However, the latter measure corresponds to
calculating the real (Schwarzschild) mass, containing a gravitational mass
defect for starlike objects with a regular center. On the other hand, in a
wormhole, there is no center at all, and $\ell =0$ corresponds to the
coordinate value $r=r_{0}$ of the throat. For the justification of using "$%
4\pi r^{2}dr$" for a centerless object, we would refer the readers to
Ref.[21a]. We thank an anonymous referee for raising this point. }:%
\begin{eqnarray}
M^{+} &=&4\pi \int_{0}^{\infty }\rho r^{2}\frac{dr}{d\ell }d\ell ,\text{ } \\
M^{-} &=&4\pi \int_{-\infty }^{0}\rho r^{2}\frac{dr}{d\ell }d\ell .
\end{eqnarray}%
As such, the integrals cannot be evaluated in a closed form although the
integrand is continuous everywhere including at $\ell =0$ and vanishing at $%
\ell \rightarrow \pm \infty $. Further, the density function $\rho
\rightarrow $ $-\frac{r_{0}^{2}}{8\pi r^{4}}$ as $\kappa \rightarrow 0$ but
it's no surprise since at this limit the EiBI theory reduces to Einstein's
theory. Also, note that $\rho \rightarrow 0$ as $\kappa \rightarrow \infty $%
. This is in perfect accordance with the pure Eddington theory ($\kappa
R\rightarrow \infty $) without matter. Thus, the behavior of $\rho $ shows
no pathology anywhere and we can legitimately expand it in powers of $\kappa 
$, which yields%
\begin{equation}
\rho =-\frac{r_{0}^{2}}{8\pi r^{4}}+\frac{3\kappa r_{0}^{4}}{16\pi r^{8}}-%
\frac{5\kappa ^{2}r_{0}^{6}}{16\pi r^{12}}+...
\end{equation}%
The limit $\kappa \rightarrow 0$ yields the first term that is just the
familiar exotic scalar field density $\rho ^{\phi }=-\frac{r_{0}^{2}}{8\pi
r^{4}}$ in the massless EB wormhole of GR. The masses can be found by term
by term integration%
\begin{eqnarray}
M^{+} &=&+\frac{r_{0}}{2}+\frac{3\kappa }{20r_{0}}-\frac{5\kappa ^{2}}{%
36r_{0}^{3}}+... \\
M^{-} &=&-\frac{r_{0}}{2}-\frac{3\kappa }{20r_{0}}+\frac{5\kappa ^{2}}{%
36r_{0}^{3}}-...
\end{eqnarray}

Note the correction terms due to $\kappa $. It is evident that the masses
are of equal value but of opposite signs. Though either mouth of the
wormhole can exhibit gravitational effects such as lensing [22] (caused
either by attractive $M^{+}$, or by repulsive $M^{-}$), the total mass of
the whole configuration adds exactly to zero, $M^{+}+M^{-}=0$, even when $%
\kappa \neq 0$. Hence, the massless character of the general relativistic EB
wormhole is preserved also in the case of its EiBI counterpart (20). In the
limit $\kappa \rightarrow 0$, one recovers the usual EB masses $+\frac{r_{0}%
}{2}$ and $-\frac{r_{0}}{2}$, which add to zero, that are made purely of the
ghost scalar field $\phi $ of GR defined by the stress $T_{\mu \nu
}=\epsilon \frac{\partial \phi }{\partial x^{\mu }}\frac{\partial \phi }{%
\partial x^{\mu }}$, with $\epsilon =-1$.

It should be noted that the Schwarzschild active gravitational masses are
trivially zero due to the fact that $g_{tt}=-1$ in the metric (20), whereas
the "bare masses" in Eqs.(24) and (25) are trivially summed to zero because
the metric is symmetric under changing $\ell \rightarrow -\ell $ and so the
derivatives $\frac{dr}{d\ell }$ in Eqs.(24) and (25) are the opposite of
each other. In the above, we showed explicit individual mass values that
could be useful for lensing purposes. 

A generalization known in GR and having nonzero masses is the twice
asymptotically flat regular massive EB wormhole [16,17,23-28], sometimes
also called the anti-Fisher solution, given by%
\begin{align}
d\tau _{\text{EB}}^{2}& =-Fdt^{2}+F^{-1}[d\ell ^{2}+(\ell
^{2}+r_{0}^{2})(d\theta ^{2}+\sin ^{2}\theta d\varphi ^{2})], \\
F& =\exp \left[ -\pi \gamma +2\gamma \tan ^{-1}\left( \frac{\ell }{r_{0}}%
\right) \right] , \\
\phi & =\lambda \left[ \pi +2\tan ^{-1}\left( \frac{\ell }{r_{0}}\right) %
\right] ,\text{ \ }
\end{align}%
with the constraint $2\lambda ^{2}=1+\gamma ^{2}.$ The Schwarzschild masses
on either side of the EB wormhole (29)-(31) are $\gamma r_{0}$ and $-\gamma
r_{0}e^{\pi \gamma }$ as can be seen by expanding the metric tensor [16,17].
Thus, when $r_{0}\neq 0$, $\gamma =0$, these masses vanish and the solution
reduces to massless EB wormhole 
\begin{eqnarray}
d\tau _{\text{EB}}^{2} &=&-dt^{2}+d\ell ^{2}+(\ell ^{2}+r_{0}^{2})(d\theta
^{2}+\sin ^{2}\theta d\varphi ^{2}), \\
\phi  &=&\frac{1}{\sqrt{2}}\left[ \pi +2\tan ^{-1}\left( \frac{\ell }{r_{0}}%
\right) \right] .
\end{eqnarray}%
Under the transformation $r^{2}=\ell ^{2}+r_{0}^{2}$, one obtains in
standard coordinates%
\begin{eqnarray}
d\tau _{\text{EB}}^{2} &=&-dt^{2}+\frac{dr^{2}}{1-r_{0}^{2}/r^{2}}%
+r^{2}[d\theta ^{2}+\sin ^{2}\theta d\varphi ^{2}], \\
\phi  &=&\frac{1}{\sqrt{2}}\left[ \pi +2\tan ^{-1}\left( \frac{\sqrt{%
r^{2}-r_{0}^{2}}}{r_{0}}\right) \right] .
\end{eqnarray}%
The metric part (\textit{sans }scalar field $\phi $) of the above solution
is the EiBI metric (20) with $\kappa =0$. As we see, it is just a special
case ($r_{0}\neq 0$, $\gamma =0$) of the metric part of the massive EB
wormhole (29)-(31).

This situation leads to a natural enquiry\footnote{%
We thank an anonymous referee for raising this query.}: Just as the metric
(20) is the EiBI generalization of the massless EB metric (34), does there
exist a similar EiBI generalization of the massive EB metric (29)? We are
not aware of such generalization as yet, but the possibility is certainly
not ruled out if, instead of the anisotropic source tensor $T^{\mu \nu }$
used by Harko \textit{et al.} [15], one uses a ghost or some other kind of
scalar field and solve the EiBI field Eqs.(8) to find a solution. This would
be a rewarding task by itself but we do not attempt it here.

\section{\textbf{Tidal forces in a Lorentz boosted frame}}

We start with the general form of a static spherically symmetric physical
metric:%
\begin{equation}
d\tau ^{2}=-\frac{F(r)}{G(r)}dt^{2}+\frac{dr^{2}}{F(r)}+R^{2}(r)[d\theta
^{2}+\sin ^{2}\theta d\varphi ^{2}].
\end{equation}%
For a traveler in a static orthonormal basis, we shall denote the only
nonvanishing components of the Riemann curvature tensor as $\mathbf{R}%
_{0101} $, $\mathbf{R}_{0202}$, $\mathbf{R}_{0303}$, $\mathbf{R}_{1212}$, $%
\mathbf{R}_{1313}$, and $\mathbf{R}_{2323}$. Radially freely falling
observers with conserved energy $E$ are connected to the static orthonormal
frame by a local Lorentz boost with an instantaneous velocity given by%
\begin{equation}
\mathbf{v}=\left[ 1-\frac{F}{GE^{2}}\right] ^{1/2}.
\end{equation}%
Then the nonvanishing Riemann curvature components in the Lorentz boosted
frame (\symbol{94}) with velocity $\mathbf{v}$ are ($k=2,3$):%
\begin{eqnarray}
\mathbf{R}_{\widehat{0}\widehat{1}\widehat{0}\widehat{1}} &=&\mathbf{R}%
_{0101} \\
\mathbf{R}_{\widehat{0}\widehat{k}\widehat{0}\widehat{k}} &=&\mathbf{R}%
_{0k0k}+\left( \mathbf{R}_{0k0k}+\mathbf{R}_{1k1k}\right) \sinh ^{2}\alpha \\
\mathbf{R}_{\widehat{1}\widehat{k}\widehat{1}\widehat{k}} &=&\mathbf{R}%
_{1k1k}+\left( \mathbf{R}_{0k0k}+\mathbf{R}_{1k1k}\right) \sinh ^{2}\alpha \\
\mathbf{R}_{\widehat{0}\widehat{k}\widehat{1}\widehat{k}} &=&\left( \mathbf{R%
}_{0k0k}+\mathbf{R}_{1k1k}\right) \sinh \alpha \cosh \alpha ,
\end{eqnarray}%
where $\sinh \alpha =\mathbf{v}/\sqrt{1-\mathbf{v}^{2}}$. The relative tidal
acceleration $\Delta a_{\widehat{j}}$ between two parts of the traveler's
body in his orthonormal basis is given by%
\begin{equation}
\Delta a_{\widehat{j}}=-\mathbf{R}_{\widehat{0}\widehat{j}\widehat{0}%
\widehat{p}}\xi ^{\widehat{p}},
\end{equation}%
where $\overrightarrow{\xi }$ is the vector separation between the two parts
[29]. Thus the curvature components contributing to tidal force on the
traveler in the Lorentz boosted frame are $\mathbf{R}_{\widehat{0}\widehat{1}%
\widehat{0}\widehat{1}}$, $\mathbf{R}_{\widehat{0}\widehat{2}\widehat{0}%
\widehat{2}}$, and $\mathbf{R}_{\widehat{0}\widehat{3}\widehat{0}\widehat{3}%
} $. (Components in the coordinate basis are not required here). In the case
of charged Reissner-Nordstr\"{o}m black hole, there occurs a remarkable
cancellation, viz., $\mathbf{R}_{0k0k}+\mathbf{R}_{1k1k}=0$ such that the
tidal accelerations in the static and moving frame are the same! The same
cancellation of course occurs in the Schwarzschild spaceime too, which is
only an uncharged special case.

For the purpose of demonstration, consider a Schwarzschild mass $M$, for
which the curvature components of interest are%
\begin{equation}
\mathbf{R}_{\widehat{0}\widehat{1}\widehat{0}\widehat{1}}=\mathbf{R}_{0101}=-%
\frac{2M}{r^{3}}\text{, }\mathbf{R}_{\widehat{0}\widehat{2}\widehat{0}%
\widehat{2}}=\mathbf{R}_{\widehat{0}\widehat{3}\widehat{0}\widehat{3}}=\frac{%
M}{r^{3}}\text{, \ etc}
\end{equation}%
Thus, at the horizon of the Schwarzschild black hole, $r_{\text{h}}=2M$, the
curvature tensor $\mathbf{R}_{\widehat{0}\widehat{j}\widehat{0}\widehat{p}%
}\varpropto \frac{1}{M^{2}}\rightarrow \infty $ as $M\rightarrow 0$. So the
smaller the black hole, the larger are the tidal forces near the horizon. We
wish to examine a similar situation near the throat of a wormhole since the
throat is physically entirely different from a black hole horizon.

\section{\textbf{Effect of }$\protect\kappa $ \textbf{on tidal forces }}

We want to calculate the effect of geodesic motion on the tidal forces
experienced by a freely falling observer. In this direction, we first note
that, in the Lorenz boosted frame, $\mathbf{R}_{\widehat{0}\widehat{1}%
\widehat{0}\widehat{1}}=\mathbf{R}_{0101}$, hence it is unaffected by
geodesic motion. Second, because of spherical symmetry, we note that $%
\mathbf{R}_{\widehat{0}\widehat{2}\widehat{0}\widehat{2}}=\mathbf{R}_{%
\widehat{0}\widehat{3}\widehat{0}\widehat{3}}$, so it is enough to calculate
only $\mathbf{R}_{\widehat{0}\widehat{2}\widehat{0}\widehat{2}}$. And
finally, with $k=2$, we can rewrite Eq.(39) for the generic metric (36) as:

\begin{equation}
\mathbf{R}_{\widehat{0}\widehat{2}\widehat{0}\widehat{2}}=-\frac{1}{R}\left[
R^{\prime \prime }\left( E^{2}G-F\right) +\frac{R^{\prime }}{2}%
(E^{2}G^{\prime }-F^{\prime })\right] ,
\end{equation}%
where primes denote derivatives with respect to $r$. The conserved energy $E$
of the falling observer can be decomposed as%
\begin{equation}
E^{2}=\frac{F}{G}+\frac{F}{G}\left( \frac{\mathbf{v}^{2}}{1-\mathbf{v}^{2}}%
\right) =E_{\text{s}}^{2}+E_{\text{ex}}^{2},
\end{equation}%
where $E_{\text{s}}^{2}$ represents the value of $E^{2}$ in the static frame
and $E_{\text{ex}}^{2}$ represents the enhancement in $E_{\text{s}}^{2}$ due
to geodesic motion. We can now decompose $\mathbf{R}_{\widehat{0}\widehat{2}%
\widehat{0}\widehat{2}}$ as follows [30]:%
\begin{eqnarray*}
\mathbf{R}_{\widehat{0}\widehat{2}\widehat{0}\widehat{2}} &=&-\frac{1}{R}%
\left[ \frac{R^{\prime }}{2}\left( E_{\text{s}}^{2}G^{\prime }-F^{\prime
}\right) \right] -\frac{1}{R}\left( R^{\prime \prime }G+\frac{R^{\prime
}G^{\prime }}{2}\right) E_{\text{ex}}^{2} \\
&=&\mathbf{R}_{\widehat{0}\widehat{2}\widehat{0}\widehat{2}}^{(\text{s})}+%
\mathbf{R}_{\widehat{0}\widehat{2}\widehat{0}\widehat{2}}^{(\text{ex})}
\end{eqnarray*}%
The first term represents the curvature component in the static frame, while
the term $\mathbf{R}_{\widehat{0}\widehat{2}\widehat{0}\widehat{2}}^{(\text{%
ex})}$ represents overall enhancement in curvature in the Lorentz-boosted
frame over that in the static frame. It is this part that needs to be
examined as the observer approaches the throat. 

Applying the above to the generalized EB wormhole metric (20), we see that%
\begin{equation}
F(r)=G(r)=\frac{1-r_{0}^{2}/r^{2}}{1+2\kappa r_{0}^{2}/r^{4}}\text{, }R(r)=r.
\end{equation}%
A little algebra will show that $\mathbf{R}_{\widehat{0}\widehat{2}\widehat{0%
}\widehat{2}}^{(\text{s})}=0$ and

\begin{equation}
\left\vert \mathbf{R}_{\widehat{0}\widehat{2}\widehat{0}\widehat{2}}^{(\text{%
ex})}\right\vert =\left[ \frac{r_{0}^{2}(r^{4}+4\kappa r^{2}-2\kappa
r_{0}^{2})}{(r^{4}+2\kappa r_{0}^{2})^{2}}\right] \left( \frac{\mathbf{v}^{2}%
}{1-\mathbf{v}^{2}}\right) ,
\end{equation}%
which, in the limit $r\rightarrow r_{0}$, gives 
\begin{equation}
\left\vert \mathbf{R}_{\widehat{0}\widehat{2}\widehat{0}\widehat{2}}^{(\text{%
ex})}\right\vert =\left( \frac{1}{2\kappa +r_{0}^{2}}\right) \left( \frac{%
\mathbf{v}^{2}}{1-\mathbf{v}^{2}}\right) .
\end{equation}%
Now suppose that\textbf{\ }$\kappa \rightarrow 0$ (general relativistic EB
wormhole) and of course $\mathbf{v\neq }$ $0$. Then, as $r_{0}\rightarrow 0$%
, the excess tidal force in the geodesic frame near the throat becomes 
\textit{arbitrarily large,} $\left\vert \mathbf{R}_{\widehat{0}\widehat{2}%
\widehat{0}\widehat{2}}^{(\text{ex})}\right\vert \rightarrow \infty $. This
behavior is very similar to, but not exactly the same as, the case of small
mass Schwarzschild black hole, as explained in Sec.4. The only physical
difference is that here we are near a narrow throat instead of a small black
hole horizon. In contrast, however, depending on the values of non-zero $%
\kappa $, the tidal forces may become \textit{arbitrarily small}, $%
\left\vert \mathbf{R}_{\widehat{0}\widehat{2}\widehat{0}\widehat{2}}^{(\text{%
ex})}\right\vert \rightarrow 0$, even when $r_{0}\rightarrow 0$. This is the
novelty of the generalized wormhole (20) brought about by the presence of
the parameter $\kappa $.

The comparison with Schwarzschild black hole as above may not be very
appropriate but still cited here only to highlight that the phenomenon of
excess curvature in the Lorentz-boosted frame was used to develop what is
called "naked black hole" in [30]. What is of interest here is the
possibility of having large or small excesses in curvature by controlling $%
\kappa $ in the generalized EB wormhole.

\section{\textbf{Flare-out and energy conditions}}

Defining $e^{-\sigma (r)}=1-\frac{m(r)}{r}$, where $m(r)$ is the
Morris-Thorne (MT) [29] shape function, and assuming that the shape of the
axially symmetric embedding surface is $z=z(r)$, the requirement that the
wormhole flares out to two asymptotically flat space times is that the
geometric condition $\frac{d^{2}r}{dz^{2}}=\frac{m-m^{\prime }r}{2m^{2}}>0$
be satisfied at or near the throat. This inequality imposes a constraint on
the type of source stress tensor $T^{\mu \nu }$, that can be nicely
rephrased in terms of the MT dimensionless flare-out function defined by: $%
\zeta :=-\frac{\rho +p_{r}}{\left\vert \rho \right\vert }=\frac{2m^{2}}{%
r\left\vert m^{\prime }\right\vert }\frac{d^{2}r}{dz^{2}}>0\Rightarrow \rho
+p_{r}<0$. It should be noted that, in spherical symmetry, the throat is
simply defined as a regular minimum areal radius and in terms of this
minimum it is easy to obtain violation of the NEC and WEC, which is really
of utmost importance but here we keep to the MT definition of flare-out.  

Harko \textit{et al.} [15] defined an alternative flare-out condition that
imposes a constraint on the shape function such that $H:=\sigma ^{\prime
}e^{-\sigma }=\frac{m^{\prime }r-m}{r^{2}}<0$ and using it obtained the
generic inequality%
\begin{equation}
8\pi \kappa (\rho +p_{r})<\frac{\kappa b^{2}}{r}\frac{\left( c^{\prime
}\right) ^{2}}{c^{2}}e^{-\sigma (r)}.
\end{equation}%
At the throat $r_{0}=m(r_{0})$, $e^{-\sigma (r)}=0$ and so $(\rho +p_{r})<0$%
. Thus, the Null Energy Condition (NEC) is violated showing that this
violation is a necessary condition for the flare-out. But if it so happens
that $\frac{\left( c^{\prime }\right) ^{2}}{c^{2}}e^{-\sigma (r)}\rightarrow
K$ as $r\rightarrow r_{0}$, then $0<\rho +p_{r}<K$, and NEC\ need not be
violated, hence no flare-out. Most importantly, note that $\rho $ and $p_{r}$
here are \textit{not} derived from the Einstein field equations using MT
metric form $e^{-\sigma (r)}=1-\frac{m(r)}{r}$, the form being used here
only for notational convenience. Instead, $\rho $ and $p_{r}$ are obtained
in Eqs.(22,23) using only the EiBI equations. Likewise, $H$ and the left
hand side of the inequality (49) are expressed in terms of the true EiBI
functions given in Sec.3.

The flare-out condition for the present wormhole (20) turns out to be%
\begin{equation}
H=\sigma ^{\prime }e^{-\sigma }=-\frac{2rr_{0}^{2}(r^{4}+4\kappa
r^{2}-2\kappa r_{0}^{2})}{(r^{4}+2\kappa r_{0}^{2})^{2}},
\end{equation}%
which, at the throat $r_{0},$ yields 
\begin{equation}
H_{0}=\left. \sigma ^{\prime }e^{-\sigma }\right\vert _{r_{0}}=-\frac{2r_{0}%
}{2\kappa +r_{0}^{2}}<0,
\end{equation}%
implying that the NEC is violated: $\rho +p_{r}<0$. We can explicitly see
from Eqs.(22) and (23) that 
\begin{equation}
\rho +p_{r}=-\frac{r_{0}^{2}}{4\pi r^{2}\sqrt{r^{4}+2\kappa r_{0}^{2}}}<0,
\end{equation}%
showing that the NEC is violated for all positive $\kappa $. The Weak Energy
Condition (WEC) is also violated for all $r$ including at the throat. As
follows from Eq.(22) 
\begin{equation}
\rho (r)=\frac{1}{8\pi \kappa }\left[ \frac{1}{\sqrt{1+2\kappa
r_{0}^{2}/r^{4}}}-1\right] <0,
\end{equation}%
for all positive $\kappa $. \textit{It is thus clear that the source of (20)
does not respect the WEC and NEC, implying that the wormhole is threaded by
exotic matter. }

It is to be noted\footnote{%
We thank another anonymous referee for pointing it out.} that $\kappa $ can
also be negative [8,12], say $\kappa =-\kappa ^{\prime }$, $\kappa ^{\prime
}>0$. Then 
\begin{equation}
\text{ }H_{0}=-\frac{2r_{0}}{r_{0}^{2}-2\kappa ^{\prime }},
\end{equation}%
which implies that $H_{0}<0$ imposes a condition on the throat radius: $%
r_{0}^{2}>2\kappa ^{\prime }$. Precisely the same condition is required for
the WEC and NEC violations as well. From Eqs.(52) and (53), we have at the
throat%
\begin{equation}
\left. \rho \right\vert _{r_{0}}=-\frac{1}{8\pi \kappa ^{\prime }}\text{ }%
\left[ -1+\frac{r_{0}}{\sqrt{r_{0}^{2}-2\kappa ^{\prime }}}\right] <0,\text{ 
}\left. (\rho +p_{r})\right\vert _{r_{0}}=-\frac{1}{4\pi r_{0}\sqrt{%
r_{0}^{2}-2\kappa ^{\prime }}}<0,
\end{equation}%
both hold only if the reality condition $r_{0}^{2}>2\kappa ^{\prime }$
holds. This suggests that the value $\sqrt{2\left\vert \kappa \right\vert }$
provides a lower bound on the size of the throat $r_{0}$, when $\kappa <0$.

Note that $H\sim ($length$)^{-1}$, while $\rho +p_{r}\sim ($length$)^{-2}$,
by definition. Hence we find a difference between the Eq.(54) and the second
of Eqs.(55), but they qualitatively mean the same physical behavior $-$
flare-out and the concomitant NEC violation respectively. The influence of $%
\kappa $ on the energy conditions and the flare-out condition is evident
from the above Eqs.(50)-(55). The individual plots of $\rho (r)$ and $%
p_{r}(r)$ exhibit similar behavior to that of $\rho +p_{r}$ and hence only
the representative plots of $\rho +p_{r}$ are given in Fig.2 for several
values of $\kappa $.

Since the wormhole (20) is threaded by exotic matter (WEC and NEC both
violated), it would be quite reasonable to enquire if EiBI exotic matter
could somehow be connected to phantom energy or ghost scalar field $\phi $
within the framework of GR. Unfortunately, this connection seems unlikely at
the level of either field equations or solutions, when $\kappa \neq 0$ (see
Appendix). The reason is that the EiBI paradigm ($\kappa \neq 0$) is very
different from that of GR ($\kappa \rightarrow 0$). Specifically, in the
EiBI field Eq.(8), the left hand side is made entirely of the auxiliary
metric $q_{\mu \nu }$, while the right hand source term $S_{\nu }^{\mu }$ is
a combination of $g_{\mu \nu }$ and $q_{\mu \nu }$ (via $\tau =\sqrt{%
\left\vert g\right\vert /\left\vert q\right\vert }$). GR limit implies
through Eqs.(6) and (7) that $\tau =1$, when $g_{\mu \nu }$ and $q_{\mu \nu
} $ become identical, and only then from Eq.(8) we end up with Einstein's
field equations.

The above notwithstanding, one might be curious to try, at the solution
level, to imbed $e^{-\sigma (r)}=1-\frac{m(r)}{r}$ and $\nu =0$ ($%
\Rightarrow $ redshift function $\Phi =0$) from (19) into the\textit{\
Einstein field equations}, and use the reverse technique of MT\ [29] to find
the GR version of the EiBI exotic matter:%
\begin{equation}
\rho _{\text{GR}}=\frac{1}{8\pi r^{2}}\frac{dm}{dr}=\frac{r_{0}^{2}\left(
6\kappa r^{2}r_{0}^{2}+4\kappa ^{2}r_{0}^{2}-6\kappa r^{4}-r^{6}\right) }{%
8\pi (r^{5}+2\kappa rr_{0}^{2})^{2}},
\end{equation}%
\begin{equation}
(\rho +p_{r})_{\text{GR}}=\frac{1}{8\pi }\left( \frac{1}{r^{2}}\frac{dm}{dr}-%
\frac{m}{r^{3}}\right) =\frac{r_{0}^{2}\left( 2\kappa r_{0}^{2}-4\kappa
r^{2}-r^{4}\right) }{4\pi (r^{4}+2\kappa r_{0}^{2})^{2}}.
\end{equation}

These are evidently very different from the corresponding EiBI Eqs.(52,53).
However, when $\kappa \rightarrow 0$, both EiBI Eq.(52) and the GR Eq.(57)
converge to the same EB value at the throat as expected, viz., $\left. (\rho
+p_{r})\right\vert _{r_{0}}=-1/4\pi r_{0}^{2}$. The plots of Eq.(57) in
Fig.3 are given for $r_{0}=1$ and several values of $\kappa $ [that is,
fixing the values of masses, see Eqs.(27,28)]. For values of $\kappa \neq 0$%
, Figs.2 and 3 exhibit different behavior. The difference is pronounced for
large negative values of $\kappa $. As an example, for $\kappa =$ $-4$,
Eqs.(56,57) give $\rho _{\text{GR}}>0$, $(\rho +p_{r})_{\text{GR}}>0$ in the
neighborhood of the throat $r\sim r_{0}=1$, i.e., no violation of WEC and
NEC, which is in contradistinction to EiBI plots in Fig.2. Nonetheless,
values of $r_{0}$ and $\kappa $ may be suitably adjusted so that $\rho _{%
\text{GR}}<0$, $(\rho +p_{r})_{\text{GR}}<0$ can also be achieved (lower
plots in Fig.3). But this GR version of EiBI exotic matter corresponds to
neither phantom nor ghost scalar field, as will be shown in the Appendix.

\section{\textbf{Light deflection}}

Light path equation in the equatorial plane, to second order in $r_{0}^{2}$,
where $\kappa $ appears first, is given by ($u=1/r$):%
\begin{equation}
\frac{d^{2}u}{d\varphi ^{2}}+u=-\left[ \frac{u}{b^{2}}+2\left( 1-\frac{%
2\kappa }{b^{2}}\right) u^{3}+6\kappa u^{5}\right] r_{0}^{2},
\end{equation}%
where $b$ is the impact parameter. The exact light path equation for the $%
\kappa =0$ case, derived earlier by Bhattacharya and Potapov [31], can be
recovered from the above. The minimum of $r$, or the maximum of $u$, denoted 
$u_{max}$, is the turning point of the motion. This occurs where $du/d\varphi
=0$ giving%
\begin{equation}
b=1/u_{\max }=R_{0}.
\end{equation}%
The perturbative solution is taken as%
\begin{equation}
u=u_{0}+u_{1}
\end{equation}%
so that the linearized equations are%
\begin{eqnarray}
\frac{d^{2}u_{0}}{d\varphi ^{2}}+u_{0} &=&0\Rightarrow u_{0}=\frac{\cos
\varphi }{R}, \\
\frac{d^{2}u_{1}}{d\varphi ^{2}}+u_{1} &=&-\left[ \frac{u_{0}}{b^{2}}%
+2\left( 1-\frac{2\kappa }{b^{2}}\right) u_{0}^{3}+6\kappa u_{0}^{5}\right]
r_{0}^{2},
\end{eqnarray}%
where $R$ is a constant. The remaining equation (62) can be integrated so
that the solution $u$ becomes: 
\begin{eqnarray}
u &=&\frac{\cos \varphi }{R}-\frac{r_{0}^{2}}{64b^{2}R^{5}}[\{56\kappa
R^{2}+16R^{4}-2b^{2}(33\kappa +14R^{2})\}\cos \varphi  \notag \\
&&+\{b^{2}(15\kappa +4R^{2})-8\kappa R^{2}\}\cos 3\varphi +b^{2}\kappa \cos
5\varphi  \notag \\
&&-(120b^{2}\kappa +48b^{2}R^{2}-96\kappa R^{2}-32R^{4})\varphi \sin \varphi
].
\end{eqnarray}%
After changing $\varphi \rightarrow \pi /2+\delta $ in Eq.(63), and assuming
small $\delta $ such that $\sin \delta \simeq \delta $, $\cos \delta \simeq
1 $, and expanding to order $r_{0}^{2}$, we find, following Bodenner and
Will [32], that%
\begin{equation}
\delta \simeq \pi r_{0}^{2}\left( \frac{3}{8R^{2}}-\frac{1}{4b^{2}}+\frac{%
15\kappa }{16R^{4}}-\frac{3\kappa }{4b^{2}R^{2}}\right) .
\end{equation}%
We now have to find the minimum value of $R$, which is the closest approach
distance $R_{0}$. The minimum of $R$ is the maximum of $u_{m}$, which can be
shown by differentiation to occur at $\varphi =0$. Putting $\varphi =0$ in
Eq.(63), setting $u_{\max }=1/R_{0}$, and inverting, we get,%
\begin{equation}
\frac{1}{R}\simeq \frac{1}{R_{0}}+O\left( \frac{1}{R_{0}^{3}}\right)
\Rightarrow R\simeq R_{0}=b.
\end{equation}%
Using this in Eq.(64), we get the two-way deflection $\epsilon $ as

\begin{equation}
\epsilon =2\delta \simeq \frac{\pi r_{0}^{2}}{4R_{0}^{2}}+\frac{3\pi \kappa
r_{0}^{2}}{8R_{0}^{4}}.
\end{equation}%
The first term\ exactly coincides with that obtained in Ref.[31], while the
second term explicitly reveals the effect of $\kappa $.

\section{\textbf{Conclusions}}

The work reported here is an extension of the work by Harko \textit{et al.}
[15], wherein they derived a wormhole solution that could be described
either as an EiBI wormhole or as generalized massless EB wormhole of GR. To
make the paper readable and understandable, we attempted to present the EiBI
basics maintaining clarity and brevity, leading the readers from the
motivation all the way to the EiBI wormhole (20) that contains a crucial
parameter $\kappa $. The value of $\kappa $ away from zero signifies
departure from general relativistic effects and has been shown in the
literature to depend on the chosen astrophysical scenarios [7-14,33,34]. In
the same spirit, we have found in the foregoing the correction terms due to $%
\kappa $ contributing to various observables in the massless EB wormhole.

We showed in Sec.3 that the massless character is preserved also in the
generalized EB wormhole (20), where $\kappa \neq 0$. In Sec.5, we found a
remarkable result is that the tidal forces can be arbitrarily small or
finite even at a small throat radius ($r_{0}\sim 0$) for \textit{non-zero}
values of $\kappa $. This result is in contradistinction to that in general
relativity, where the tidal forces become arbitrarily large in the limit of
small Schwarzschild horizon radius ($M\sim 0$), as argued in the previous
Sec.4. Then we discussed in Sec.6 the inter-relations among $\kappa $, the
flare-out and energy conditions in EiBI showing that the source of (20) does
not respect the WEC and NEC for $\kappa >0$. For $\kappa <0$, the throat
radius has a lower bound $2\sqrt{\left\vert \kappa \right\vert }$ for $\rho $
and $\rho +p_{r}$ to be real, but the energy conditions are still not
respected. Posing the EiBI wormhole as a general relativistic one, we find
that energy conditions may or may not be respected depending on the choices
of $r_{0}$ and $\kappa $. This is more of a curious GR exercise as we show
in the Appendix that the EiBI wormhole cannot be fitted into the GR
framework with a phantom or ghost source scalar field $\phi $ even with a
potential $V(\phi )$. In Sec.7, we have shown that the two-way light
deflection on the positive side of the mouth has a correction term
proportional to $\kappa $.

Some immediate tasks remain: The gravitational lensing by the general
relativistic ($\kappa =0$) EB wormhole has been already investigated by Abe
[22]. Hence it would be of interest to study the influence of $\kappa \neq 0$
on the lensing observables in the generalized metric (20) taking into
account our correction term to light deflection obtained in Eq.(66). Another
important question is the issue of stability. It is already shown within the
framework of GR that the $\kappa =0$ case is unstable both under linear and
non-linear perturbations [23,24,25] only if the EB wormhole has a phantom
scalar as a source. The same metric can be obtained with another source, an
exotic fluid, then the dynamics is quite different, and the equation of
state of this fluid can be chosen in such a way that this wormhole will be
stable. All this is explicitly shown in [35]. Stability of the generalized
wormhole (20) has to be studied within the framework of EiBI theory for
which $\kappa \neq 0$ and it is yet to be understood if the presence of
non-zero $\kappa $ can allow stability.

\textbf{Acknowledgments}

This work was supported in part by an internal grant of M. Akmullah Bashkir
State Pedagogical University in the field of natural sciences. We thank two
anonymous referees for their insightful comments that have led to a
considerable improvement of the paper.

\section{\textbf{References}}

[1] A. Riess \textit{et al.}, Astron. J. \textbf{116}, 1009 (1998)

[2] S. Perlmutter \textit{et al.}, Astrophys. J. \textbf{517}, 565 (1999)

[3] S. M. Carroll \textit{et al.}, Phys. Rev. D \textbf{70}, 043528 (2004)

[4] S. Deser and G.W. Gibbons, Class. Quant. Grav.\textbf{15}, L35 (1998)

[5] M. Ba\~{n}ados and P. G. Ferreira, Phys. Rev. Lett. \textbf{105}, 011101
(2010)

[6] M. Born and L. Infeld, Proc. Roy. Soc. A \textbf{144}, 425 (1934)

[7] J. Casanellas, P. Pani, I. Lopes and V. Cardoso, Astrophys. J. \textbf{%
745}, 15 (2012)

[8] P. Pani, V. Cardoso and T. Delsate, Phys. Rev. Lett. \textbf{107},
031101 (2011)

[9] T. Harko \textit{et al}., Phys. Rev. D \textbf{88}, 044032 (2013)

[10] J. H. C. Scargill, M. Ba\~{n}ados, and P. G. Ferreira, Phys. Rev. D 
\textbf{86}, 103533 (2012)

[11] P. P. Avelino and R. Z. Ferreira, Phys. Rev. D \textbf{86}, 041501(R)
(2012)

[12] P. Pani and T.P. Sotiriou, Phys. Rev. Lett. \textbf{109}, 251102 (2012)

[13] Q.-M. Fu \textit{et al}., Phys.Rev. D \textbf{90}, 104007 (2014)

[14] X.-L. Du \textit{et al.}, Phys. Rev. D \textbf{90}, 044054 (2014)

[15] T. Harko \textit{et al}., Mod. Phys. Lett. A \textbf{30}, 1550190 (2015)

[16] H.G. Ellis, J. Math. Phys. \textbf{14}, 104 (1973); \textit{Errata}: J.
Math. Phys. \textbf{15}, 520 (1974)

[17] K.A. Bronnikov, Acta Phys. Polon. B \textbf{4}, 251 (1973)

[18] S.D. Odintsov \textit{et al.}, Phys. Rev. D \textbf{90}, 044003 (2014);
M. Lagos \textit{et al.}, Phys. Rev. D \textbf{89}, 024034 (2014); 
T. Harko \textit{et al}., Galaxies \textbf{2}, 496 (2014); 
A.N. Makarenko \textit{et al.}, Phys. Rev. \textbf{90}, 024066 (2014); 
M. Bouhmadi-L\'{o}pez \textit{et al}., JCAP 11 (2014) 007; 
M. Bouhmadi-L\'{o}pez \textit{et al}., Phys.Rev. D \textbf{90}, 123518 (2014); 
M. Bouhmadi-L\'{o}pez \textit{et al}., Eur. Phys. J. C \textbf{75}, 90 (2015); 
S.-W. Wei \textit{et al.}, Eur. Phys. J. C  \textbf{75253} (2015); 
R. Shaikh, hys. Rev. D \textbf{92}, 024015 (2015); 
E. Berti \textit{et al}., arXiv:1501.07274 [gr-qc]

[19]  A.S. Eddington, \textit{The Mathematical Theory of Relativity (Cambridge University Press, Cambridge, UK, 1924)}.

[20] D. N. Vollick, Phys. Rev. D \textbf{69} (2004) 064030;\textit{\ ibid.} D%
\textbf{\ 72}, 084026 (2005)

[21] M. Visser, S. Kar and N. Dadhich, Phys. Rev. Lett.\textbf{\ 90}, 201102
(2003); S. Kar, N. Dadhich and M. Visser, Pramana,\textbf{\ 63}, 859 (2004)

[22] F. Abe, Astrophys. J. \textbf{725}, 787 (2010)

[23] J.A. Gonz\'{a}lez, F.S. Guzm\'{a}n and O. Sarbach, Class. Quant. Grav.%
\textbf{\ 26}, 015010 (2009); \textit{ibid}. \textbf{26}, 015011 (2009)

[24] K.A. Bronnikov, J.C. Fabris and A. Zhidenko, Euro. Phys. J. C \textbf{71%
}, 1791 (2011)

[25] K. A. Bronnikov, R. A. Konoplya and A. Zhidenko, Phys. Rev. D \textbf{86%
}, 024028 (2012)

[26] K.A. Bronnikov and S. Grinyok, Grav. Cosmol. \textbf{10}, 237 (2004)

[28] K.A. Bronnikov and A.A. Starobinsky, JETP Lett. \textbf{85},1 (2007)

[29] M.S. Morris and K.S. Thorne, Am. J. Phys. \textbf{56}, 395 (1988)

[30] G.T. Horowitz and S.F. Ross, Phys. Rev. D \textbf{56}, 2180 (1997)

[31] A. Bhattacharya and A. A. Potapov, Mod. Phys. Lett. A \textbf{29}, 2399
(2010)

[32] J. Bodenner and C.M. Will, Am. J. Phys. \textbf{71}, 770 (2003)

[33] R. Izmailov \textit{et al}., Mod. Phys. Lett. A \textbf{30}, 1550056
(2015)

[34] A.A. Potapov \textit{et al}., JCAP 07 (2015) 018

[35] K.A. Bronnikov, L.N. Lipatova, I.D. Novikov and A.A. Shatskiy, Grav.
Cosmol. \textbf{19}, 269 (2013) 

\begin{center}
\textbf{Appendix}
\end{center}

We shall show that the exotic matter threading the EiBI or generalized EB
wormhole (20) ($\kappa \neq 0$) is neither phantom nor ghost in the GR
framework. For phantom matter, the equation of state parameter should be $%
\omega =$ $\frac{p_{r}}{\rho }<-1$. On the other hand, we have from
Eqs.(22,23) 
\begin{equation}
\omega =\frac{p_{r}}{\rho }=\sqrt{1+2\kappa r_{0}^{2}/r^{4}}>0,\forall
\kappa ,r  \tag{A1}
\end{equation}%
including at the throat $r=r_{0}$. The EiBI exotic matter therefore cannot
be phantom anywhere in the spacetime regardless of whether $\kappa $ is
positive or negative.

However, for the $\kappa =0$, it is well known that the EB wormhole (34) is
threaded by matter made purely of a minimally coupled ghost scalar field in
GR. The question then we ask is: Can we find in GR a similar minimally
coupled scalar field $\phi $ with an arbitrary potential $V(\phi )$ for the $%
\kappa \neq 0$ EiBI wormhole (20)? The answer, unfortunately, seems to be in
the negative.

Consider the action with a minimally coupled scalar field $\phi $ and a
potential $V(\phi )$ given by 
\begin{equation}
S=\frac{1}{8\pi }\int d^{4}x\sqrt{-g}\left[ R-\epsilon (\nabla \phi
)^{2}-2V(\phi )\right] ,  \tag{A2}
\end{equation}%
where, notationally, $(\nabla \phi )^{2}\equiv g^{\mu \nu }\phi _{\mu }\phi
_{\nu }$, $\phi _{\mu }\equiv \partial \phi /\partial x^{\mu }$ and $%
\epsilon =\pm 1$. Variation with respect to the metric $g_{\mu \nu }$ and $%
\phi $ gives respectively the field equations%

\begin{equation}
R_{{\mu}{\nu }}=\epsilon \phi _{\mu }\phi _{\nu }+g_{\mu \nu }V,  \tag{A3}
\end{equation}

\begin{equation}
\phi _{;\alpha }^{;\alpha }=-\frac{\partial V}{\partial \phi }. \tag{A4}
\end{equation}

The value $\epsilon =-1$ corresponds to what is called a ghost scalar field $%
\phi $. We choose the metric ansatz%
\begin{equation}
d\tau ^{2}=-B(r)dt^{2}+A(r)dr^{2}+r^{2}[d\theta ^{2}+\sin ^{2}\theta
d\varphi ^{2}].  \tag{A5}
\end{equation}%
From the left hand side of the field equation (A3), since $\overset{.}{A}=0$%
, it follows that $R_{tr}=R_{rt}=\frac{\overset{.}{A}}{2A}=0$, so we get $%
\overset{.}{\phi }\phi ^{\prime }=0$, where prime denotes differentiation
with respect to $r$ and dot denotes differentiation with respect to $t$. So
we can either have $\phi ^{\prime }=0$ or $\overset{.}{\phi }=0$. We choose
the latter and assume $\phi =\phi (r)$ so that we get from the Eqs.(A3):%

\begin{equation}
\frac{B^{\prime \prime }}{2A}-\frac{B^{\prime }}{4A}\left( \frac{A^{\prime}}{A}+\frac{B^{\prime }}
{B}\right) +\frac{B^{\prime }}{rA} =V,  \tag{A6}
\end{equation}

\begin{equation}
-\frac{B^{\prime \prime }}{2B}+\frac{B^{\prime }}{4B}\left( \frac{A^{\prime }%
}{A}+\frac{B^{\prime }}{B}\right) +\frac{A^{\prime }}{rA} =\epsilon \phi
^{\prime 2}+AV,  \tag{A7} 
\end{equation}

\begin{equation}
1-\frac{1}{A}+\frac{rA^{\prime }}{2A^{2}}-\frac{rB^{\prime }}{2AB} =r^{2}V.
\tag{A8}
\end{equation}

For the EiBI metric (20), we have%
\begin{equation}
B(r)=1,\text{ }A(r)=\frac{1+2\kappa r_{0}^{2}/r^{4}}{1-r_{0}^{2}/r^{2}}. 
\tag{A9}
\end{equation}%
Putting them in (A6), we have $V=0$ but the difficulty is that the field
equation (A8), viz., 
\begin{equation}
1-\frac{1}{A}+\frac{rA^{\prime }}{2A^{2}}=0  \tag{A10}
\end{equation}%
is \textit{not} satisfied by the function $A(r)$ unless $\kappa =0$. This
lack of self-consistency indicates that the exotic source matter in (20) is
unlikely to be represented by a GR ghost scalar field. Note that although
the GR Eqs.(56,57) yield (for suitable values of $r_{0}$ and $\kappa $)
exotic source matter obtained via the reverse MT [29] method, unless we are
able to derive them from some kind of exotic scalar field via action of the
type (A2), we cannot connect the solution (20) with $\kappa \neq 0$ to a GR
solution with a coupled scalar field $\phi $ typical of the EB solutions.

However, there is always the possibility to introduce ghost or phantom or
some other scalar field into the EiBI theory\textit{\ itself} by including
them in the action (1) from the start and analyze the corresponding
solutions, if any. That would be a separate task by itself and is left for
the future. Having said that, we point out that Deser and Gibbons [4]
considered the EiBI type of Lagrangian and took the usual Christoffel
connection $\Gamma (g)$ [instead of $\Gamma (q)$] and treated $g_{\mu \nu }$
as the only dynamical variable. The resulting field equations were fourth
order with ghosts [20]. But the EB solutions result from second order field
equations with ghost source, and thus different from the one considered in
[4].

\begin{center}
\textbf{Figure captions}
\end{center}

\begin{figure} 
\includegraphics [width=\linewidth] {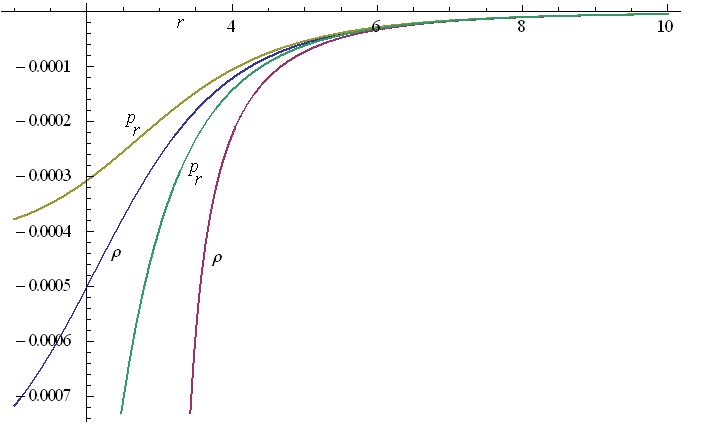}
\caption{Plot of $\rho $ and $p_{r}$ vs $r$ of EiBI Eqs.(22,23) at $r_{0}=1$.
The red and blue curves for $\rho $ correspond to $\kappa =-50$ and $50$
respectively, while the green and grey curves for $p_{r}$ correspond to $%
\kappa =-50$ and $50$ respectively. The values of $\rho $ and $p_{r}$ are
always negative for arbitrary values of $r_{0}$ and $\kappa $. Only some
representative plots are displayed.}
\end{figure}

\begin{figure}
\includegraphics [width=\linewidth] {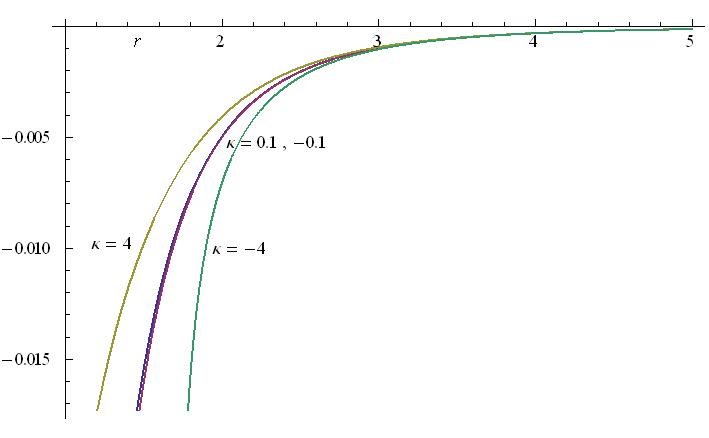}
\caption{Plot of $\rho +p_{r}$ vs $r$ of EiBI Eq.(52) at $r_{0}=1$. It shows
that NEC is violated for positive and negative values of $\kappa $. Similar
curves follow for arbitrary values of $r_{0}$ and $\kappa $. Only some
representative plots are displayed.}
\end{figure}

\begin{figure}
\includegraphics [width=\linewidth] {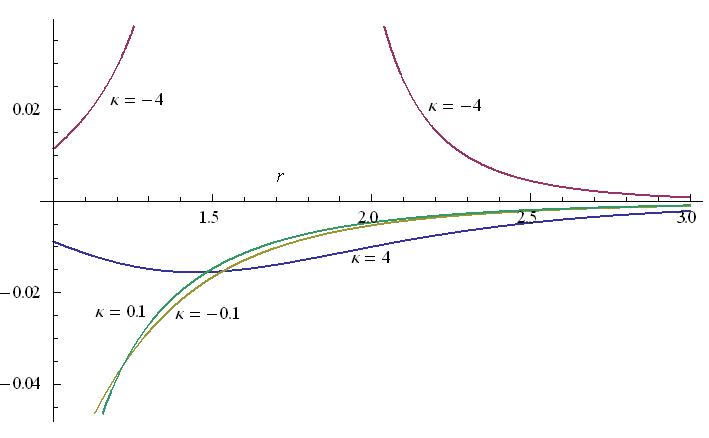}
\caption{Plot of $\rho +p_{r}$ vs $r$ of GR Eq.(57) at $r_{0}=1$. For
relatively large negative $\kappa $, say, $\kappa =-4$, NEC is not violated.
Values of $r_{0}$ and $\kappa $ can be adjusted to have NEC violation. Only
some representative plots are displayed.}
\end{figure}

\end{document}